\documentclass[conference]{IEEEtran}

\usepackage[latin1]{inputenc}
\usepackage{amsmath,amsfonts,amssymb}
\usepackage{stmaryrd} 
\usepackage{tabularx}
\usepackage{graphicx}
\usepackage{epstopdf}
\usepackage{enumerate}
\usepackage[parfill]{parskip}

\usepackage{arydshln} 

\usepackage{xkeyval,calc,listings,tikz}

\newtheorem{thm}{Theorem}[section]

\newtheorem{dfn}{Definition}[section]

\newtheorem{corollary}{Corollary}

\begin{document}
\title{Coding Bounds for Multiple Phased-Burst Correction and Single Burst Correction Codes}

\author{\IEEEauthorblockN{Wai Han Fong}\IEEEauthorblockA{Department of Electrical and Computer Engineering\\George Mason University\\Fairfax, Virginia  22030, USA\\Email: wfong@masonlive.gmu.edu}} 

\maketitle

\begin{abstract}
In this paper, two upper bounds on the achievable code rate of linear block codes for multiple phased-burst correction (MPBC) are presented.  One bound is constrained to a maximum correctable cyclic burst length within every subblock, or equivalently a constraint on the minimum error free length or gap within every phased-burst.  This bound, when reduced to the special case of a bound for single burst correction (SBC), is shown to be the Abramson bound when the cyclic burst length is less than half the block length.  The second MPBC bound is developed without the minimum error free gap constraint and is used as a comparison to the first bound.
\end{abstract}
\begin{IEEEkeywords}
Multiple phased-burst correction bounds, burst correction bounds, coding bound, code rate bounds, cyclic burst error.
\end{IEEEkeywords}

\section{Introduction}

\IEEEPARstart{W}ireless channels must contend with error inducing phenomena that cause multiple burst errors. Data storage devices also suffer from errors that can also occur in multiple bursts. A solution to these problems can be found from multiple burst correction coding.  This technique has been studied previously, \cite{mandelbaum2002some}, \cite{zhi2002constructions}, \cite{stone1961multiple} and \cite{bridwell1970burst}. Furthermore, methods for correcting bursts that are confined to subblocks (\emph{phased-bursts}), or multiple phased-burst correction (MPBC) coding, were studied in \cite{goodman2002phased}, \cite{blaum2002new} and \cite{keren2002codes}. 

The goal of this paper is to provide a measure of the code rate efficiency of MPBC codes.  Two upper bounds on the achievable MPBC code rate are presented where one of the bounds can be used as a single burst correction (SBC) bound as a special case. The derivation for this MPBC bound is based on an unambiguous definition of a cyclic burst error. This definition allows us to analyze burst error correction codes when the phase-burst length is greater than half of the subblock length for MPBC or analogously when the cyclic burst length is greater than half the codelength for SBC.  For SBC, this has been a difficult problem to solve in the past \cite{reiger1960codes}, \cite{fire1959class}, \cite{camp1962}, \cite[p. 202] {mceliece2002theory}, which have led these authors to derive bounds with burst error lengths that are constrained to be less than half the codelength. This problem is solved here by counting burst error patterns with maximum zero strings that are less than the error free length or gap of the error pattern.  The error free gap can be considered the minimum error free space that separates multiple bursts. It is shown that when the MPBC bound is reduced to a SBC bound, it becomes a generalization of the Abramson bound which is a corollary of the Hamming bound for burst-error correction \cite[p. 202] {mceliece2002theory}.  The ability to enumerate burst error patterns based on the specification of maximum zero strings within the burst error of a certain weight, sets this work apart from \cite{sharma2002extended} which looked at SBC bounds for burst errors of a maximum weight.

\section{Preliminaries}

A discussion of the background concepts of end-around intervals and burst error patterns is presented.  This is followed by a discussion of phase-bursts and MPBC codes.  

An \emph{end-around} interval of length $l$, where $0\le l < n$ of a binary vector of length $n$, is an interval that starts at position $l_{begin}$ and ends at position $l_{end}=(l_{begin}+l)_n-1$, where $(\cdot)_n$ denotes the modulo $n$ operation and $l_{begin},l_{end}\in\{0,1,\dots,n-1\}$.  Furthermore, if $l$ is large enough, then $l_{begin} > l_{end}$.  An \emph{error pattern} is a binary vector of length $n$ where the non-zeros are the locations of symbol errors.  A \emph{burst error} is an error pattern of length $n$ where the symbol errors are localized in an interval of length $l$ where the first and last positions of the burst are non-zeros.  In \cite[p. 200]{mceliece2002theory}, a \emph{cyclic burst} is defined as a burst error where the location of the burst is an end-around interval of length $l_{burst}$.  This definition, however, is not free of ambiguity.  As noted in \cite[p. 200]{mceliece2002theory}, the starting position of the cyclic burst could be at a number of non-zero positions, each with different burst lengths.  In order to avoid this ambiguity, a constraint which is called in this paper, the \emph{cyclic burst constraint} (CBC), is defined to constrain the burst length $l_{burst}$ \cite[p. 201]{mceliece2002theory}: 
\begin{equation}\label{eqn:cbc}
l_{burst}\le \lfloor(n+1)/2\rfloor.
\end{equation}
Equation (\ref{eqn:cbc}), can also be interpreted as lower bound on the end-around \emph{error free space} $l_{error\:free}$ surrounding a cyclic burst:
\begin{equation}\label{eqn:fspace}
l_{error\:free}=n-l_{burst}>n-\lfloor(n+1)/2\rfloor=\lfloor(n-1)/2\rfloor.
\end{equation}
The CBC allowed for the unambiguous analysis of burst correction coding, however, SBC coding bounds adhered strictly to the CBC \cite{reiger1960codes}, \cite{fire1959class}, \cite{camp1962}, \cite[p. 202] {mceliece2002theory}.

Now consider a binary $(n,k)$ linear block code $\cal{C}$ that has codewords $\bf c$ which can be partitioned into $t$ subblocks of length $v$ symbols, i.e. 
\[
{\bf c}=[{\bf\underline c}_0, {\bf \underline c}_1, \ldots, {\bf\underline c}_{t-1}],
\]
is composed of $t$ phases of length $v$ or $n=tv$, where ${\underline  {\bf c}}_j=[c_{jv}, c_{jv+1}, \ldots, c_{jv+v-1}]$ and $0\leq j<t$, is the $j$th \emph{codeword-phase}.  
A \emph{phased-burst} is a burst error of length $u$ confined in a subblock  or codeword-phase in ${\bf c}$.  Block codes with this codeword structure that can correct multiple phased-bursts are called MPBC codes \cite[p. 1118]{LinCostello_ErrorControlCoding_2004}.

\section{Multiple phased-burst error correction bounds}

In this section, two MPBC bounds are developed.  One based a precise definition of a cyclic phased-burst error (or with regard to an error free gap) and another bound based on burst errors that have a maximum number of correctable symbols per subblock (i.e. no gap constraint). Another interpretation of an error free gap is that if error bursts are not confined to separate codeword-phases then a burst can crossover a subblock boundary.  In this case, multiple burst errors must separated by a minimum error free gap otherwise there is no distinction between a large burst and multiple smaller bursts. 

First multiple-burst correction codes with cyclic phased-burst errors are considered with the goal to develop a cyclic phased-burst MPBC bound based on an unambiguous definition of a cyclic burst. To do this, the CBC is a consideration since it is applicable to every subblock because the MPBC bound can be an SBC bound as a special case.  That is, since (\ref{eqn:fspace}) is a bound on the minimum error free space of an SBC code, it can also be seen as a bound on the minimum error free gap of a phased-burst and if not mitigated, the CBC will constrain the minimum error free gap to be greater than half the length of a subblock.  However, in order to remove the CBC from the MPBC bound, a new cyclic burst definition must be provided that can be unambiguously applied.  

\begin{dfn}\label{dfn:err_burst} 
An \emph{end-around phased-burst error} of length $u$ in a subblock of length $v$ contains no consecutive string of zeros of length $g=v-u$ or more within a burst error pattern of length $u$ that is an end-around sequence. 
\end{dfn}

This definition specifies that a minimum end-around guard space or gap $g$ be maintained within a subblock.  This guard space is by definition error free, and to avoid ambiguity, no other string of zeros within the burst can be equal to or greater than $g$. As an example, (\ref{eqn:burst_pattern})
\begin{equation}\label{eqn:burst_pattern}
\begin{split}
&\leftarrow\!\!\!|\:\:\:\:\:\:\:\:\:\:\:\:\,|\!\!\!\rightarrow \\       
0&1010000010001010001\\
\end{split}
\end{equation}
 shows a $v=20$ burst error pattern that is indexed left-to-right from 0 to 19 with the largest string of zeros of length 5 starting at position 4 and ending at position 8. This defines the error free gap $g=5$.  An end-around error burst of length $u=15$ starts at position 9 and ends at position 3 as indicated by the arrows.  Within the burst there are zero strings of length 3 and 1 but none that are equal to or greater than $g$.  

There are two consequences of this definition when the error free gap $g\le \lfloor(v-1)/2\rfloor$, i.e. when the burst does not conform to the CBC. The first consequence is that there are possible error patterns where the largest string of zeros occur multiple times.  This is interpreted as a multiple burst condition within a subblock which is not considered in calculating the bound. The second consequence is that when the CBC is not conformed to, the burst length will be larger than the gap and creates a lower bound on the number of ones in the burst or the \emph{burst weight} $w_{burst}>2$ according to the theorem below (proof omitted): 

\begin{thm}\label{thm:bweight}
Let an end-around phased-burst error pattern of length $v$ have a burst of length $u$ and an error free gap of $g=v-u$, then the \emph{burst weight} is bounded by $w_{burst}\ge \lceil \frac{u-1}{g} \rceil +1$.
\end{thm}

From the above theorem, it's clear that when the $g \ge u-1$, i.e. conforms to the CBC, the \emph{minimum burst weight} $w_{burst,\:min}=2$.   This is the case where the burst only consists of one string of zeros bounded by two non-zeros.  However, when $0<g < u-1$, then $3\le w_{burst,\:min} \le u$.  Thus, the region where the burst becomes larger than the error free gap, is also the region where the minimum burst weight increases above 2.  As seen below, this increase in $w_{burst,\:min}$, will maintain a high upper bound on the achievable code rate of an MPBC code.  

After specifying an unambiguous definition of a cyclic burst and exploring its ramifications, the MPBC bound is now developed. From coding theory, a linear block code is capable of correcting the set of all error patterns that are defined as coset leaders. The approach in crafting a bound is to enumerate all possible cosets leaders that conforms to Definition \ref{dfn:err_burst} in all subblocks.  The goal is to be able to count all binary patterns of a certain length based on specifying the largest string of zeros in a pattern given a specification for the total number of non-zeros in the pattern.  This result can be used to guarantee that no patterns of zero strings are larger than the gap specification.  The following theorem provides the means to enumerate these patterns.

\begin{thm}\label{thm:1st_bound_thm}
Let ${A}(c,d,e)$ be the number of non-zero binary patterns of length $c$ with the number of ones $d$, that has a maximum consecutive string of zeros of $e$ or less. Then the number of non-zero binary vectors ${B}(x,y,z)$ of length $x$ with the number of ones $y$, that has a maximum consecutive string of zeros of $z$ is: 
\begin{equation}\label{eqn:theta}
{B}(x,y,z)={A}(x,y,z)-{A}(x,y,z-1)
\end{equation}
where $x-(y+z) \ge 0$ and
\begin{equation}\label{eqn:beta}
{A}(c,d,e)=\sum_{j\in\mathbb{J}}(-1)^j{{d+1}\choose{j}}{{c-j(e+1)}\choose{d}} \end{equation}
\end{thm}
where $\mathbb{J}=\{j:0\le j\le d+1,c-j(e+1)\ge0\}$.

\begin{IEEEproof}
See Appendix \ref{app1}.
\end{IEEEproof}

From Theorem \ref{thm:1st_bound_thm}, the enumeration of all patterns based on a maximum zero string length is possible.  Since a burst is bounded by two non-zeros, (\ref{eqn:theta}) can be used to count the possible patterns that occurs between the two non-zero boundary symbols. And in order to maintain Definition \ref{dfn:err_burst}, patterns within a subblock that have zero strings equal to or larger than the gap $g$ are not allowed.  Theorem \ref{thm:1st_bound_thm} provides the ability to enumerate patterns within the interval between the two non-zero boundary symbols of a burst based on the largest zero strings as a parameter.  The MPBC bound can now be calculated by the following theorem.

\begin{thm}\label{thm:2nd_bound_thm}
Let $F(x,y,z,v)$ be the number of binary vectors of a subblock of length $v$ with a burst error of length $x+2$, that has $y+2$ number of non-zeros and a maximum zero strings of length $z$.  Then an $M$ multiple phase correcting linear block code of length $n=tv$ and dimension $k$, where $v$ is the length of each subblock, $t$ is number of subblocks in a codeword, $M$ is maximum number of correctable subblocks, and $u$ is the maximum length of a correctable cyclic phased-burst per subblock according to Definition \ref{dfn:err_burst}, has the number of coset leaders $2^{n-k}$ bounded by:
\begin{equation}\label{eqn:Mbound}
2^{n-k}\ge\sum_{j=1}^{M}{{t}\choose{j}}\left[\sum_{x=0}^{u-2}\sum_{y=0}^{x}\sum_{z=0}^{v-x-3}{{v}\choose{1}}F(x,y,z,v)\right]^j+n+1
\end{equation}
where 
\begin{equation}\label{eqn:xi}
F(x,y,z,v)=
\begin{cases}
1, & (y=0) \wedge (x=z < \lfloor \frac{v}{2}-2 \rfloor),\\
{B}(x,y,z), & otherwise.
\end{cases}
\end{equation}
\end{thm}

\begin{IEEEproof}
See Appendix \ref{app2}.
\end{IEEEproof}
Equation (\ref{eqn:Mbound}) can be restated in terms of code rate $r_c=\frac{k}{n}$:  
\begin{equation}\label{eqn:MeRbound2}
\begin{split}
r_c \le 1-\frac{1}{n}\log_2 (\sum_{j=1}^{M}&{{t}\choose{j}}\left[\sum_{x=0}^{u-2}\sum_{y=0}^{x}\sum_{z=0}^{v-x-3}{{v}\choose{1}}F(x,y,z,v)\right]^j \\ 
& +n+1).
\end{split}
\end{equation}
The case without a gap constraint is now considered. In this instance, the MBPC codes are limited by a specified maximum number of correctable symbols per subblock and the maximum number of correctable subblocks.  This bound, presented in the theorem below (proof omitted), is used to compare the impact to the achievable code rate when the error free gap constraint is observed.

\begin{thm}\label{thm:3nd_bound_thm}
The number of coset leaders, $2^{n-k}$, of a multiple phase correction linear block code of length $n=vt$ and dimension $k$, where  $v$ is the length of each subblock, $t$ is number of subblocks in a codeword,  $M$ is maximum number of correctable subblocks with a maximum number of correctable symbols per subblock $E$ is bounded by:
\begin{equation}\label{eqn:Mebound}
2^{n-k}\ge\sum_{j=1}^{M}{{t}\choose{j}}\left[\sum_{l=1}^{E}{{v}\choose{l}}\right]^j+1.
\end{equation}
\end{thm}

Equation (\ref{eqn:Mebound}) can also be restated in terms of code rate: 
\begin{equation}\label{eqn:MeRbound}
r_c\le1-\frac{\log_2(\sum_{j=1}^{M}{{t}\choose{j}}\left[\sum_{l=1}^{E}{{v}\choose{l}}\right]^j+1)}{n}.
\end{equation}
and for $M=1$ and $t=1$, reduces to the well known Hamming bound for binary codes.

\begin{figure}[htbp] 
   \centering
   \includegraphics[width=3.5in]{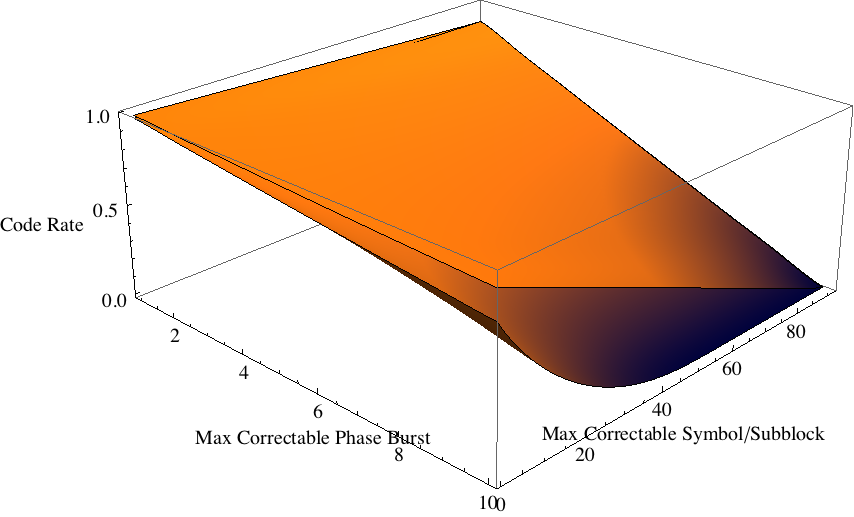} 
   \caption{MPBC Bounds with and without a Gap Constraint}
   \label{fig:mecombbound}
\end{figure}

To show the differences between the two bounds, a plot of them together is shown in Fig. \ref{fig:mecombbound} for a code consisting of 10 subblocks and a subblock length of 100 bits.  The gap constrained bound (\ref{eqn:MeRbound2}) is on top while the bound without the gap constraint (\ref{eqn:MeRbound}) is below.  The maximum number of correctable phased-burst $M$ ranges from 1 to 10, while the maximum correctable symbols/subblock, $u$ in (\ref{eqn:MeRbound2}) and $E$ in (\ref{eqn:MeRbound}), ranges from 5 to 95.  The achievable region for either bounds is the region below the surface, respectively.  This shows that the gap constrained bound, whose surface is flat, is higher in achievable code rate than the bound without the gap constraint, whose surface is concave.  The difference between the two surfaces increases as the maximum number of correctable phased-bursts increases.  For a fixed value of maximum number of correctable phased-bursts, the difference between the two bounds will be nearly zero while approaching the edges at 5 and at 95 maximum correctable symbols/subblock. The difference will increase as the maximum correctable symbols/subblock tends toward mid-range values and reaches a peak at a maximum correctable symbol/subblock of 34 for a particular $M$. The largest difference in maximum achievable code rate is 0.502 occurring at the $M=10$ and a maximum correctable symbol/subblock of 34. 

As a special case, where the entire subblock are correctable, i.e. $E=v$, the right side of (\ref{eqn:MeRbound}) is reduced to:
\begin{equation}\label{eqn:Mebound2}
2^{n-k}\ge\sum_{j=1}^{M}{{t}\choose{j}}(2^{v}-1)^j+1
\end{equation}
The example in Fig. \ref{fig:mecombbound} showed that the two bounds would approach the same value as the maximum correctable symbols/subblock approach the length of the subblock, i.e. $E=v$. 
In this instance, there is no gap constraint and burst errors are corrected to the entire subblock length. In this case, when (\ref{eqn:MeRbound2}) is set to its smallest possible gap number, i.e. $g=1$, it will approach the same result as (\ref{eqn:Mebound2}).  

\section{Single burst correction bound}

In this section, the MPBC bound (\ref{eqn:Mbound}) is considered under at a special case of $M=1$ and $t=1$ as a upper bound on the achievable code rate of an SBC code.  To do this, Theorem \ref{thm:2nd_bound_thm} is reduced to this corollary (proof omitted): 

\begin{corollary}\label{cor:singleburst}
A single burst correction code with block code of length $n$, dimension $k$ and maximum correction burst length $u$ has the minimum number of coset leaders $2^{n-k}$ bounded by:
\begin{equation}\label{eqn:Sbound}
2^{n-k}\ge{{n}\choose{1}}\sum_{x=0}^{u-2}\sum_{y=0}^{x}\sum_{z=0}^{n-x-3}F(x,y,z,n)+n+1
\end{equation}
\end{corollary}

Equation (\ref{eqn:Sbound}) can be restated in terms of code rate:
\begin{equation}\label{eqn:Sbound3}
r_c\le1-\frac{\log_2({{n}\choose{1}}\sum_{x=0}^{u-2}\sum_{y=0}^{x}\sum_{z=0}^{n-x-3}F(x,y,z,n)+n+1)}{n}
\end{equation} 

Corollary \ref{cor:singleburst} in the form of (\ref{eqn:Sbound3}) gives an SBC upper bound on the achievable code rate that is not constrained to the CBC. To explore the connections with previously published bounds, the following corollary is used (proof omitted):  

\begin{corollary}\label{col:conject}
For a burst error pattern of length $n$ whose burst length $x+2$ is constrained under the CBC, the double summation of $F(x,y,z,n)$ over variables $0 \le y \le x$ and $0 \le z \le x$ is equal to all possible binary vectors of length $x$: 
\begin{equation}\label{eqn:conject}
\sum_{y=0}^{x}\sum_{z=0}^{x}F(x,y,z,n)=2^x.
\end{equation}
\end{corollary}

Corollary \ref{col:conject} can be used to evaluate (\ref{eqn:Sbound}) under the CBC condition.  
Then (\ref{eqn:Sbound}) becomes:
\begin{equation}\label{eqn:Sbound2}
2^{n-k}\ge{{n}\choose{1}}\sum_{x=0}^{u-2}\sum_{y=0}^{x}\sum_{z=0}^{x}F(x,y,z,n)+n+1.
\end{equation}
That leads to:
\begin{equation}\label{eqn:Sshortbound}
2^{n-k}\ge n\sum_{x=0}^{u-2}2^x+n+1.
\end{equation}
 The summation is a geometric series that will equal to $2^{u-1}-1$ and therefore (\ref{eqn:Sshortbound}) becomes  
\begin{equation}\label{eqn:Sshortbound2}
2^{n-k}\ge n(2^{u-1}-1)+n+1=n2^{u-1}+1
\end{equation}

Under the CBC condition, equation (\ref{eqn:Sshortbound2}) is precisely the Hamming bound for burst-error correction and when written in terms of required parity bits $n-k$ becomes the Abramson bound  \cite[p. 202]{mceliece2002theory}. And thus, (\ref{eqn:Sbound}) is a generalization of the Abramson bound without the CBC on the burst length.

\section{Conclusion}

Two MPBC bounds on the maximum achievable code rate have been presented.  One MPBC bound is based on an unambiguous definition of a cyclic burst error which allows burst lengths to exceed the CBC and under special case of SBC is proven to be a generalization of the Abramson bound.  The example results show that cyclic phased-burst MPBC codes have a high achievable code rate due to the error free gap constraint.  This fact is made clear when compared to the MPBC bound without the gap constraint with the difference between the bounds increasing as the maximum number of correctable phased-bursts increases.

\appendices
\section{Proof of Theorem \ref{thm:1st_bound_thm}}\label{app1}
Equation (\ref{eqn:beta}) is proved first.  In combinatorics, the Sieve Theorem \cite[p. 47]{balakrishnan1995schaum} can be stated as follows: let $\mathbb{X}$ be a finite set and have subsets $\mathbb{U}_i$ where $1\le i \le L$, then 
\begin{equation}\label{eqn:sieve}
\arrowvert\bigcap_{i=1}^{L} \mathbb{U}^C_i\arrowvert=|\mathbb{X}|-\sum_{j=1}^{L}(-1)^js_j
\end{equation}
where $s_j$ denotes the sum of the cardinalities of all the $j$-tuple intersections of the $L$ subsets $\mathbb{U}_i$, $\mathbb{U}^C_i$ is the complement of $\mathbb{U}_i$ and $1\le j \le L$.  To find the number of patterns where the all zero strings are less than or equal to length $e$, the Sieve Theorem (in a similiar approach but for a different application as that found in \cite[Prob. 2.21, pp. 54-55]{balakrishnan1995schaum}) is used to find the intersection of events of zero strings greater than $e$ for any possible $d+1$ positions. 

Let $\mathbb{X}$ be defined as the set of all patterns of length $c$ with $d$ non-zeros and therefore: $|\mathbb{X}|={{c}\choose{d}}$.  And let $\mathbb{U}^C_i$ be defined as the event that the length of the zero string at $i$ is less than or equal to $e$, where $1\le i \le L=d+1$.  Then $\mathbb{U}_i$ is the event that the length of the zero string at $i$ is greater than or equal to $e+1$ and the cardinality of the intersection of all $\mathbb{U}^C_i$ is given by (\ref{eqn:sieve}).  The next step is to find a general equation for $s_j$, which is defined as the cardinality of all possible intersection of $j$ $\mathbb{U}_i$ events.  

In composition theory, the number of integer solutions of the equation $\sum t_i=t_{total}$ for which every $t_i\ge l_i$ and $\sum l_i=l_{total}$ where $1\le i \le r$ is equal to ${{t_{total}-l_{total}+{r}-1}\choose{{r}-1}}$ \cite[Prob. 1.142, p. 36]{balakrishnan1995schaum}.  If $t_i$ represents the length of the zero string at position $i$ where $1 \le i \le r=d+1$, then $\sum t_i=t_{total}$ is the total number of zeros in the pattern which is also the length of the pattern minus the number of ones, i.e. $t_{total}=c-d$.  And if the constraint that $t_i$ be greater than or equal to $l_i=e+1$ for a subset of $j$ positions, i.e.
\begin{equation}
l_i=
\begin{cases}
e +1, & 1\le i \le j\\
0, & j+1\le i \le d+1
\end{cases}
\end{equation}
 then $l_{total}=j(e+1)$.  Therefore the number of patterns for $j$ positions of zero strings of length greater than $e$ is ${{c-j(e+1)}\choose{d}}$.  Since there are ${{d+1}\choose{j}}$ possible combinations of selecting $j$ positions from $d+1$ positions, $s_j={{d+1}\choose{j}}{{c-j(e+1)}\choose{d}}$.    Then from (\ref{eqn:sieve}), all patterns with zero strings of length $e$ or less is $\sum_{j=0}^{d+1}(-1)^j{{d+1}\choose{j}}{{c-j(e+1)}\choose{d}}$ where the $|\mathbb{X}|$ term is incorporated into summation for $j=0$.  However the last binomial coefficient term can be undefined if $c-j(e+1)<0$, therefore the summation is limited accordingly by defining the set $\mathbb{J}=\{j:0\le j\le d+1,c-j(e+1)\ge0\}$ to get (\ref{eqn:beta}). 
 
To find the total number of patterns with a maximum zero string of length $z$, the total number of patterns of maximim zero strings of $z-1$ or less are subtracted from the total number of patterns of maximum zero strings of $z$ or less to get (\ref{eqn:theta}). This equation is valid only for $x-(y+z) \ge 0$ since all calculations must involve non-negative lengths. $\hfill\blacksquare$ 

\section{Proof to Theorem \ref{thm:2nd_bound_thm}}\label{app2}
Based on coding theory, for a given linear block code there are $2^{n-k}$ coset leaders that are correctable error patterns. By counting all patterns that follow Definition \ref{dfn:err_burst} for a specific minimum gap distance and subblock length, all possible error patterns for one subblock using Theorem \ref{thm:1st_bound_thm} to lower bound the number of coset leaders are enumerated.  

Given a subblock of length $v$, all error patterns must be have an error free gap $g \ge v-u$ to correct a burst of length $u$ or less.  Since any burst including end-around bursts must be bounded by non-zeros, Theorem \ref{thm:1st_bound_thm} can be used to calculate patterns of length $u-2$ or smaller.  If $b=x+2$ specifies the length of a burst under consideration, then $x$ must vary from $0\le x \le u-2$ so that bursts of length $2\le b \le u$ are considered.  In every case, the burst can only contain zero strings that are less than the current error free gap under consideration, i.e. $0 \le z< v-(x+2)$.  Under these conditions, ${B}(x,y,z)$ calculates all except the all zeros case which can be accounted for by defining the function $F(x,y,z,v)$.  According to Theorem \ref{thm:bweight}, the all zero case occurs when the burst length conforms to the CBC, i.e. minimum burst weight of 2 since two non-zeros bounds a burst.  Therefore, $F(x,y,z,v)=1$ only when the following events intersect: 1) the number of zeros is the same as the pattern length, i.e. $z=x$; 2) under CBC conformance: $g = v-(x+2) > x+2$ then $x< \lfloor \frac{v}{2}-2 \rfloor$; and 3) that there are no non-zeros, $y=0$.   By applying (\ref{eqn:xi}), the first parameter $x$ defines a pattern length that must start from zero and ends at a value less than or equal to $u-2$.   $y$ starts from zero to the pattern length $x$. $z$ starts from zero and is limited by the constraint of being smaller than the current gap, i.e. $v-x-3$. Summing over all cases would give all possible error patterns given a particular placement of gap $g$ in a subblock of length $v$.  Since there are ${{v}\choose{1}}$ possible locations for the start of the gap, the previous calculation is multiplied by ${{v}\choose{1}}$. In this way, the end-around burst will be accounted for. This calculation is the total number of end-around patterns that are correctable within a subblock for a given minimum gap $g$. If there are $j$ correctable subblocks, then this result needs to be raised to the $j^{th}$ power since the patterns in each subblock are disjoint.  If the largest number of subblocks that need to be corrected is $M$, then this calculation is multiplied by the number of possible combinations for every number of correctable subblocks up to $M$ and summed, i.e. a partial sum of binomial coefficients of ${{t}\choose{j}}$, where $1\le j \le M$. Finally, $n+1$ is added to account for $n$ single bit error patterns and $1$ for the all zero pattern.  The result is ($\ref{eqn:Mbound}$).  $\hfill\blacksquare$

\bibliographystyle{IEEE}
\bibliography{burst_Ref}

\end{document}